\documentclass[preprint,aps,10pt,twocolumn]{revtex4}
\usepackage{graphicx}% Include figure files
\usepackage{dcolumn}% Align table columns on decimal point
\usepackage{amsmath}
\usepackage{bm}% bold math
\usepackage{natbib}
\usepackage[mathlines]{lineno}%
\graphicspath{{figs/}} % Directory in which figures are stored
\begin{document}
%\preprint{APS/123-QED}
\title [Application of....] {Application of high frequency biasing and its effect in STOR-M tokamak}
\author{Debjyoti Basu$^{a,b}$, Masaru Nakajima$^{a}$, A.V. Melnikov$^{c,d}$, Julio J. Martinell$^{e}$, David McColl$^{a}$,  Raj Singh$^{b}$, Chijin Xiao$^{a}$, Akira Hirose$^{a}$}
\address{$^{a}$Plasma Physics Laboratory, University of Saskatchewan, Saskatoon, Canada}
\address{$^{b}$ Present address: Institute for Plasma Research, Bhat, Gandhinagar-382428, India}
\email{debjyotibasu.basu@gmail.com}
\address{$^{c}$ NRC Kurchatov Institute, 123182, Moscow, Russia}
\address{$^{d}$ National Research Nuclear University MEPhI, 115409, Moscow, Russia}
\address{$^{e}$ Instituto de Ciencias Nucleares-UNAM, Mexico D.F. 04510, Mexico}
\email{martinel@nucleares.unam.mx}
\date{\today}% It is always \today, today,

\begin{abstract}
A pulsed oscillating power amplifier has been developed to apply high frequency biasing voltage to an electrode at the edge of the STOR-M tokamak plasma. The power amplifier can deliver a peak-to-peak oscillating voltage up to 120V and current 30A within the frequency range of 1kHz-50kHz. The electrode is located in the equatorial plane at radius $\rho = 0.88$. The frequency of the applied voltage has been varied between discharges. It is observed that the plasma density and soft x-ray intensity from the plasma core region usually increase at lower frequency regime 1kHz-5kHz as well as relatively higher frequency regime 20kHz-25kHz but seldom increase in between them. Increment of $\tau_{p}$ \& $\tau_{E}$ have been observed from the derivations of experimental data in both frequency regimes. Transport simulation has been carried out using the ASTRA simulation code for STOR-M tokamak parameters to understand the physical process behind experimental observations at higher frequency branch. The model is based on GAM excitement at resonance frequency associated with Ware-pinch due to oscillating electric field produced by biasing voltage which can suppress anomalous transport. Simulation results reproduce the experiment quite well in terms of the density, particle confinement as well as energy confinement time evolution. All those results indicate high frequency biasing is capable of improving confinement efficiently.
\end{abstract}
%Valid PACS numbers may be entered using the \verb+\pacs{#1}+ command.

\pacs{Valid PACS appear here}% PACS, the Physics and Astronomy
%%                             % Classification Scheme.
\maketitle
%\twocolumn
\section{Introduction}
Edge plasma turbulence in tokamak plays a crucial role in plasma transport and confinement \cite{kn:vanoost1}. In general, it is believed that anomalous transport in tokamaks is governed by plasma turbulence. Experimentally, it has been established that anomalous transport happens mostly due to edge electrostatic turbulence \cite{kn:Ritz,kn:Liewer}, but the basic mechanism of edge turbulence is not clear till now. Another interesting feature of turbulence phenomena is the possibility to drive zonal flows and GAM \cite{kn:Diamond, kn:deb1} when it reaches certain energy level and may helps to trigger H-mode in tokamak. Turbulence has interesting nature and great importance related to tokamak plasma confinement where it generally worsens plasma confinement but sometimes deterioration is reduced, leading to a relatively better plasma confinement. Therefore, high frequency biasing \cite{kn:shur} may be another approach for understanding the role of turbulence in plasma confinement.
\par Previously, series of dc and low frequency (around few hertz) ac electrode biasing \cite{kn:Taylor, kn:Weynants, kn:deb, kn:Melnikov, kn:Xiao, kn:silva} and turbulent feedback experiments \cite{kn:Kan, kn:UCKAN, kn:Brooks} have been performed to understand the nature of turbulent transport and improved confinement achieved by edge turbulence suppression. Recently, it was planned to perform high frequency (kilo-Hertz range), electrode biasing experiments on the STOR-M tokamak to study dependency of transport on biasing voltage frequency. So, an ac broadband (1kHz-50kHz) power amplifier has been developed to meet experimental needs to amplify broadband signal without phase shift.
\par In this paper, a brief description of power supply development, experimental results and analysis of its physical nature through transport simulation using ASTRA code will be discussed.
\par The STOR-M tokamak is a limiter based small tokamak with circular plasma cross-section having major and minor radii 46 cm and 12 cm respectively. In high-frequency bias experiments, an oscillating voltage of $\pm 60$V has been applied by the developed power supply at plasma edge through an electrode made of rectangular stainless steel plate located in equatorial plane at plasma radius $10.5(\rho=0.88)$ cm. Frequency of applied voltage has been varied from $1$kHz to $25$kHz on shot to shot basis. A 12 channel pin-hole soft x-ray camera masked by a 1.8$\mu$m aluminum foil and viewed from top port\cite{kn:Xiao1} has been used to measure line-integrated soft x-ray emission intensity. A half-meter monochromator for $H_{\alpha}$ recording and microwave interferometer for line averaged electron density measurement have been used.
\begin{figure}[h]
\center
\includegraphics[width=260pt,height=230pt]{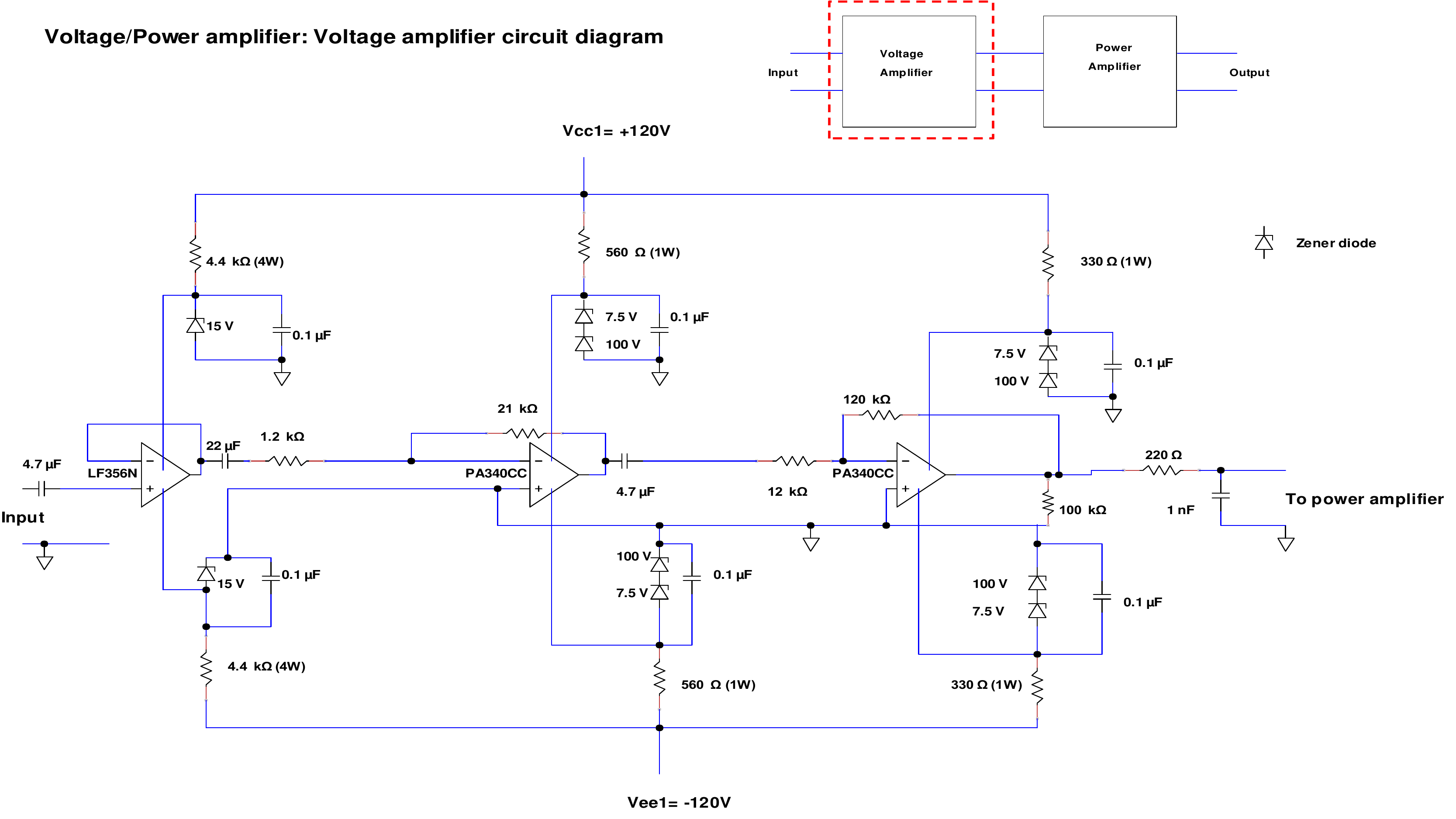}
\caption{Voltage amplifier section.} \label{fig:1}
\end{figure}
\begin{figure}[h]
\center
\includegraphics[width=260pt,height=230pt]{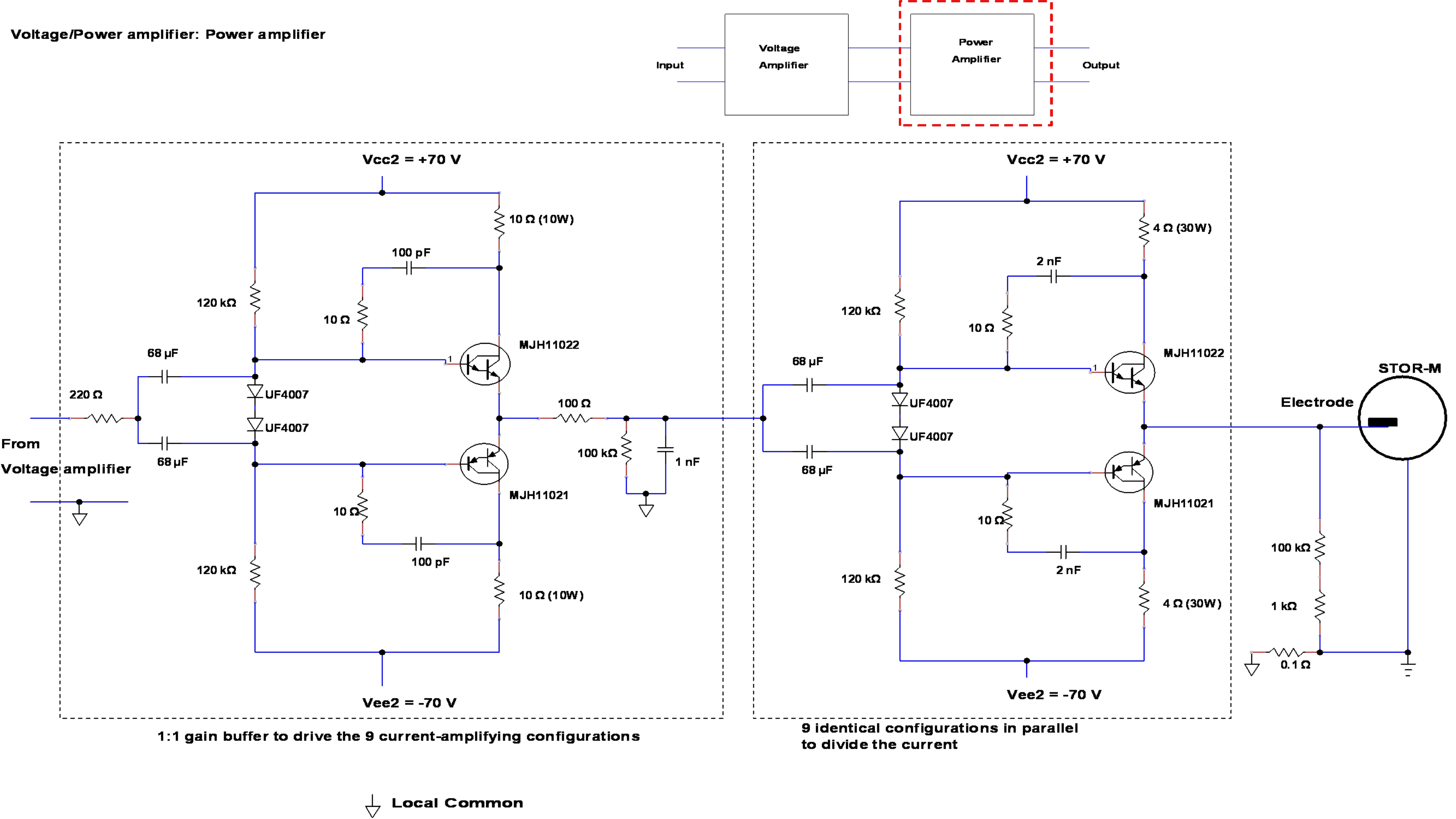}
\caption{Current amplifier section.} \label{fig:2}
\end{figure}
%\begin{figure}[ht]
%\center
%\includegraphics[width=240pt,height=200pt]{fig1.pdf}
%\caption{High frequency biasing flowchart diagram.}\label{fig:3}
%\end{figure}

\par The power amplifier can deliver output signal with peak voltage $\pm60$V and current $30$A without any distortion in any waveform when its input voltage is $0.25$V and current $0.5$A. The selected frequency range has been chosen based on the dominant frequency of the drift mode driven by density gradient present in STOR-M. The power amplifier has two key parts which amplify voltage(figure 1) \& current(figure 2) and thus the overall power of waveform. Voltage amplification has been done through high frequency and power mosfet op-amp PA340CC. Voltage has been amplified by two inverting op-amp with amplification factor around $175$. MJH11021 PNP and MJH11022 NPN darlington pairs with ratings  $250$V \& $10$A have been used for current amplification. Nine identical push-pull amplifiers are connected in parallel to deliver the combined high power required for the experiments. A voltage divider and 0.1$\Omega$ sampling resistor are used to monitor electrode voltage and current through plasma. The power amplifier is gated and driven by a function generator. Activation and operational time duration of amplifier is controlled by pulsed controller circuit using solid state switch LF13202. The pulsed controller circuit is turned on by a master optical trigger pulse. Pulse delay and its width can be varied from 0.5ms-44ms and 1ms-45 ms, respectively.
\par Frequency was varied between shots from 1kHz to 25kHz. Interestingly, biasing improves confinement for frequency regimes of 1kHz-5kHz \& 20kHz-25kHz. The confinement was rarely improved outside these two frequency regimes. A typical plasma shot with and without biasing is shown in figure 3.
\begin{figure}[ht]
\center
\includegraphics[width=220pt,height=180pt]{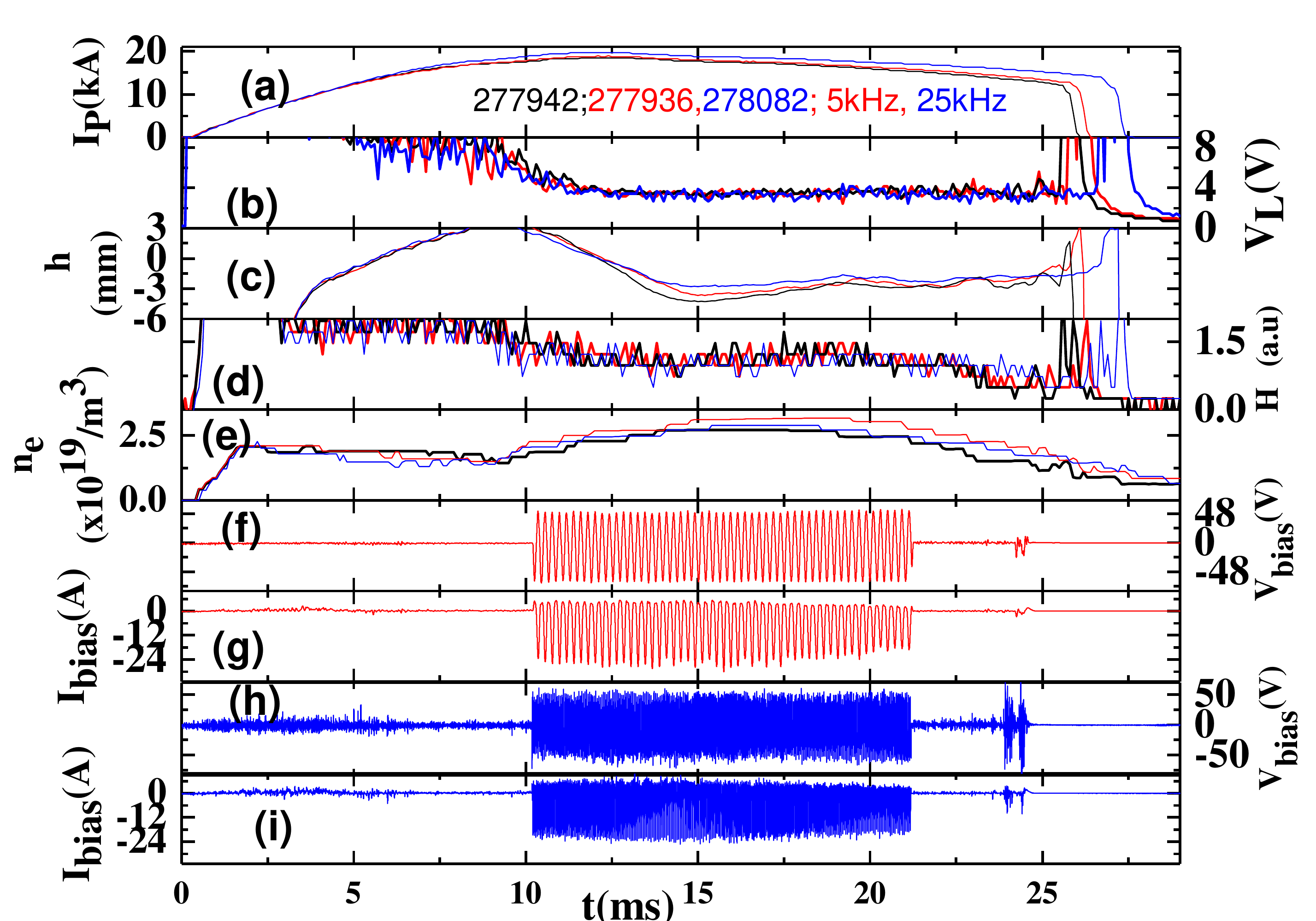}
\caption{Temporal evolution of plasma parameters without (black line) \& with (red line 5kHz \& blue line 22kHz) biasing (a) plasma current, (b) loop voltage, (c) plasma horizontal position, (d) Plasma $H_{\alpha}$ intensity, (e) line averaged density, (f) bias voltage, (g) bias current at low frequency; (h) bias voltage, (i) bias current at high frequency.}\label{fig:3}
\end{figure}
\par In this typical example, a triangular voltage signal with peak to peak voltage of $\pm60$V was applied from 10ms to 21ms at 5kHz \& 25kHz during the plateau of Tokamak discharge current as shown in figure 3. Here, an important feature is that  overall plasma density has increased without significant change in $H_{\alpha}$ intensity, shown in figure 3(d) \& 3(e). Horizontal plasma equilibrium position started to shift during application of high frequency bias voltage but came back to its equilibrium position due to position feedback control, shown in figure 3(c). Interestingly, figure 3(f), 3(g), 3(h) \& 3(i) show that maximal value of the electrode current decreases with time while the electrode voltage remains unchanged indicating improved confinement\cite{kn:Taylor,kn:Heikkinen}.
\par Time evolution of central soft x-ray channel and particle confinement time $(\tau_{p})$ with and without biasing are presented in figure 4(a) \& 4(b). $\tau_{p}$ is the ratio of electron density $n_{e}$ to $H_{\alpha}$ intensity\cite{kn:Hidalgo}, which has enhanced due to biasing. Increment of soft x-ray signals from central and its adjacent channels has been noticed from radial profile in presence of biasing in time window 15.5ms-16.5ms, shown in figure 4(c).
\begin{figure}[ht]
\center
\includegraphics[width=240pt,height=200pt]{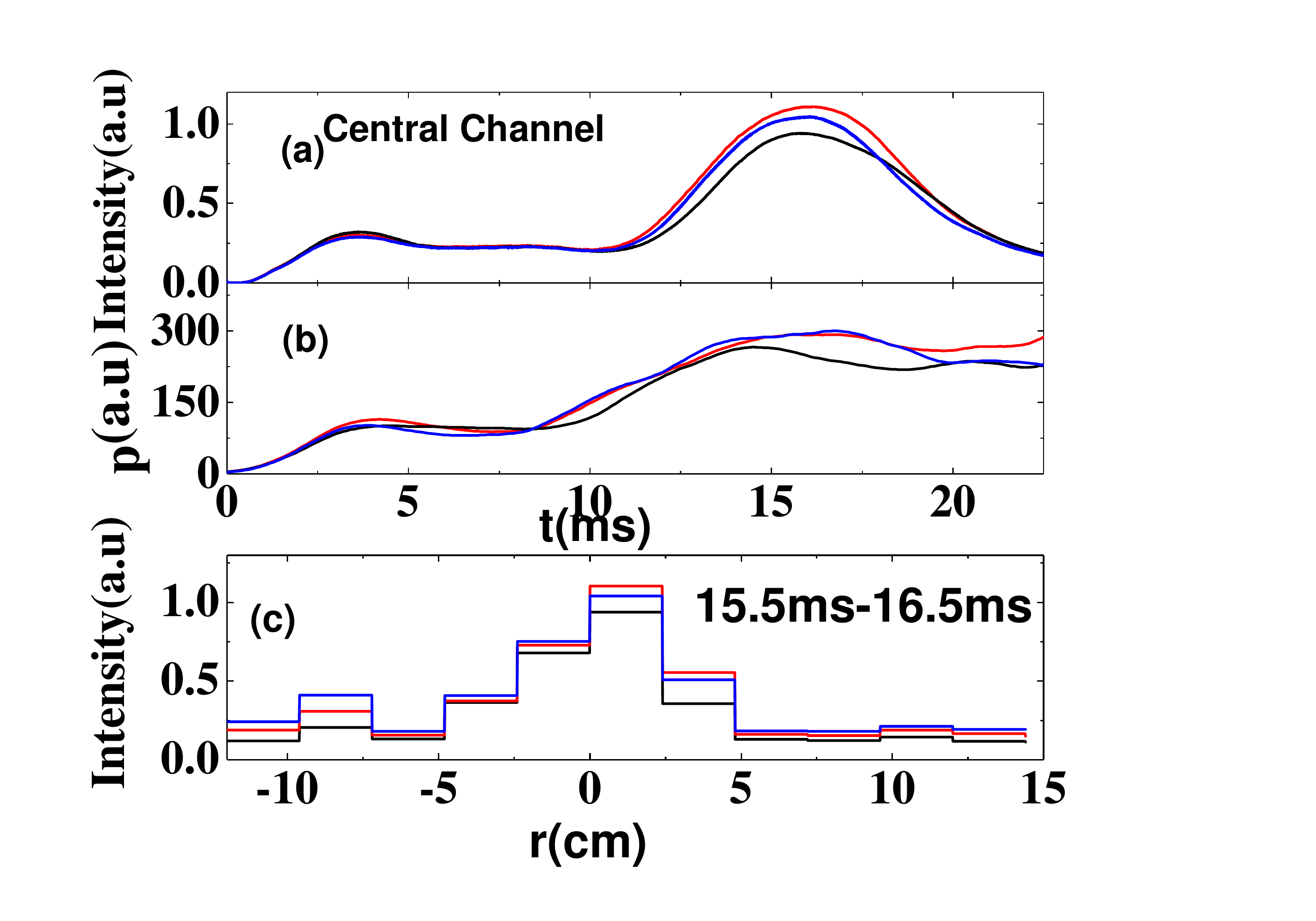}
\caption{Temporal evolution of (a) soft x-ray central channel, (b)representation of particle confinement time, (c)radial profile of soft x-rays in between $15.5ms-16.5ms$ with(red line 5kHz \& blue line 22kHz) \& without(black line) bias.}\label{fig:4}
\end{figure}
\par These results indicate improvement of core plasma confinement may happen due to biasing. The nature of density variation and soft x-ray behaviour at different frequencies are shown in figure 5. Interestingly, it is observed in figure 5(a) \& 5(b) that changes are sometimes opposite in soft x-ray and density profile with different bias frequencies. If density is lower at lower frequency then soft x-ray intensity is higher at lower frequency compared to that at higher frequency. In this typical example, density and soft x-ray profiles at 3kHz, 20kHz \& no bias(0kHz) have been compared. To quantify this, percentage increment of $n_{e}.T_{e}$(a.u) has been derived within 15.5ms-18.5ms at lower frequencies (1kHz-5kHz) as well as higher frequencies (20kHz-25kHz) where maximum change happened for both profiles. Soft x-ray intensity due to Bremsstrahlung has been used for derivation of percentage increment of $n_{e}.T_{e}$(a.u). The radiation due to bremsstrahlung which is received by soft x-ray detectors can be written as $I_{br}=cn_{e}^{2}T_{e}^{1/2}Z_{eff}$ \cite{kn:silver}, where, $I_{br}$, $n_{e}$, $T_{e}$ are in $watt/m^{3}$,$ m^{-3}$, $eV$ respectively and `$Z_{eff}$' is effective ion charge state. Here, it is considered that `$Z_{eff}$' remains unaltered in both situations for different frequencies because comparison between with \& without bias cases show that $I_{p}$ is not decreased and $V_{L}$ \& $H_{\alpha}$ are not increased. 
\begin{figure}[ht]
\center
\includegraphics[width=240pt,height=200pt]{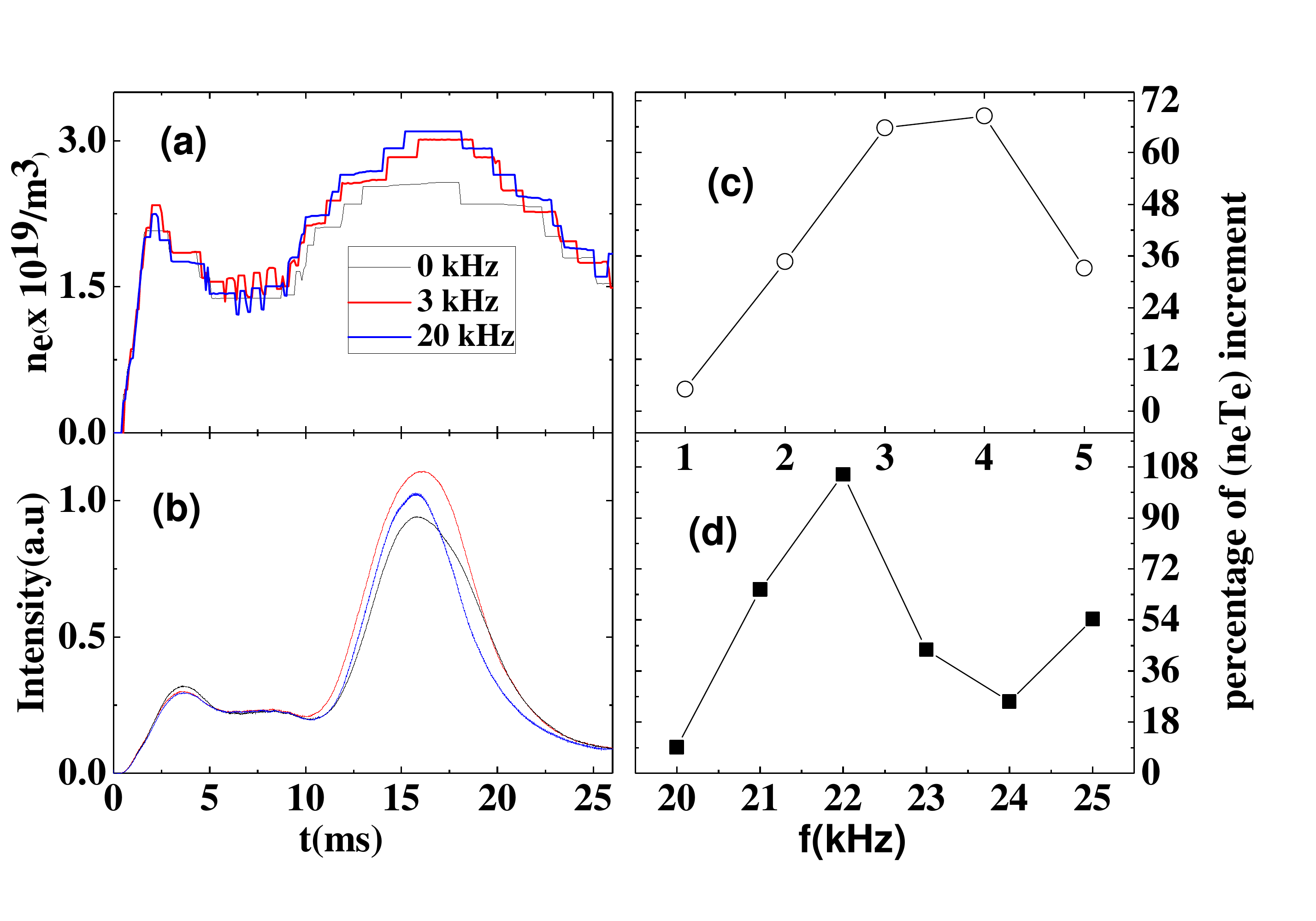}
\caption{Temporal evolution of (a) density, (b) soft x-ray central channel at no bias (black), 3kHz(red), 20kHz (blue), frequency dependance of percentage increment of $n_{e}. T_{e}$ at (c) lower frequency branch, (d) higher frequency branch.}\label{fig:5}
\end{figure}
\par Smoothed signals of $V_{L}$ \& $H_{\alpha}$ using adjacent averaging of 25 points have been plotted in figure 6(a) \& 6(b) within time window of 12ms-24ms. $V_{L}$ and $H_{\alpha}$ profile within 15.5ms-18.5ms clearly show that they are decreased with biasing. Intensity coming from same soft x-ray channel has been compared for with and without bias cases. Since, soft x-ray comes from around center, density value from interferometer data is used for getting a crude approximation. Here, percentage increment of a physical quantity `$\alpha$' is defined as $\frac{\alpha_{with bias}- \alpha_{without bias}}{\alpha_{without bias}}\times 100 $. Energy percentage increment reaches maximum at 4kHz \& 22kHz as shown in figure 5(c) \& 5(d). The increment of $n_{e}T_{e}$ is indication of plasma beta enhancement which is supported by Grad-Shafranov shifting of plasma column horizontally from its equilibrium position as shown in figure 6(c).
\begin{figure}[ht]
\center
\includegraphics[width=240pt,height=200pt]{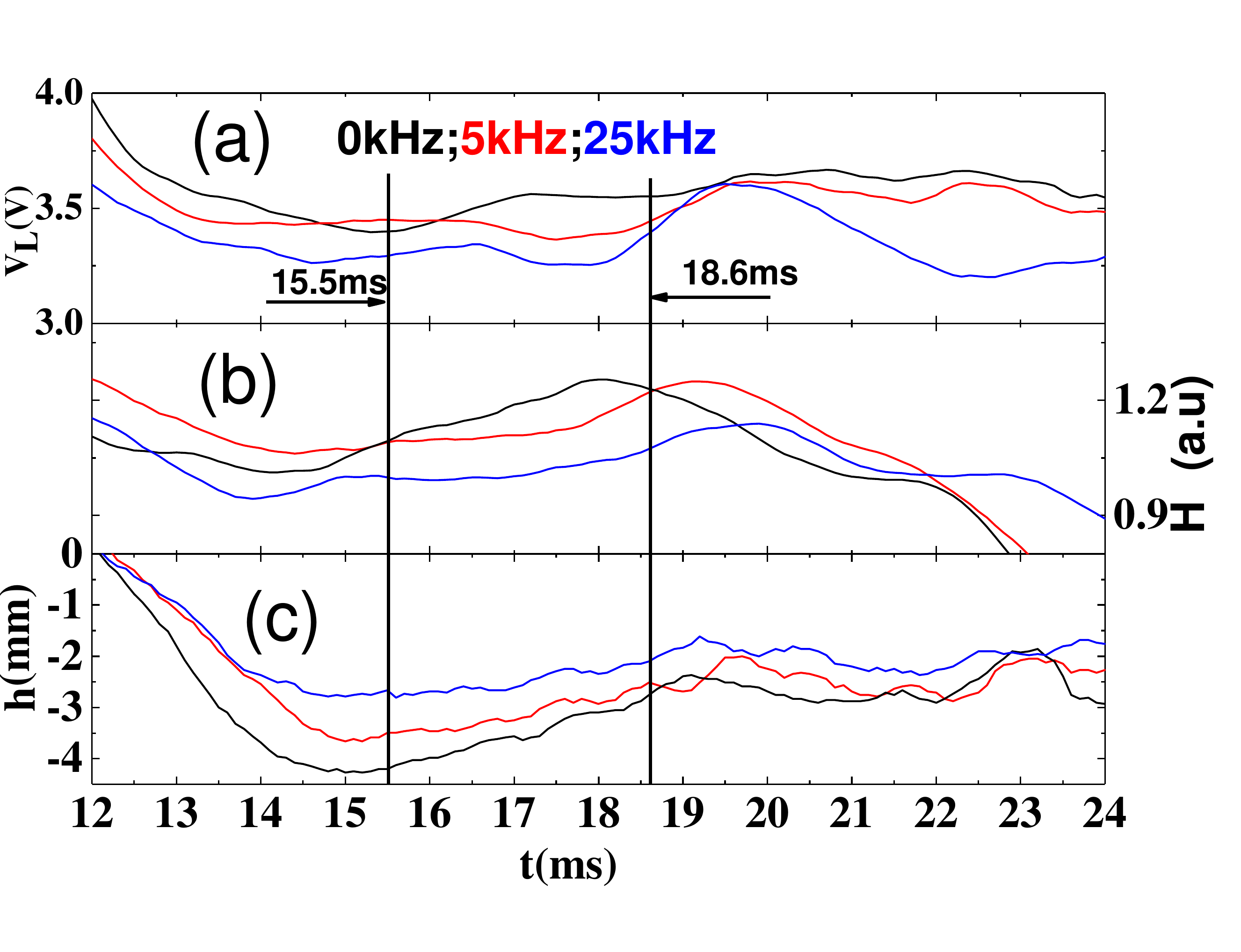}
\caption{Temporal evolution of zoomed (a) loop voltage (smoothed 25 points), (b) $H_{\alpha}$-intensity (smoothed 25 points), (c)horizontal position $(\Delta_{h})$ at no bias (black), 5kHz(red), 25kHz (blue).}\label{fig:6}
\end{figure}
 \par Percentage increment of $\tau_{p}$ \& a new physical quantity $\tau_{E\_}I_{soft}$ are shown in figure 7(a)\& 7(b)for lower and higher frequency regimes respectively, where $\tau_{E\_}I_{soft}\propto \frac{I_{br}^{2}}{n_{e}^{3}I_{p}V_{L}}$ and $\tau_{p} \propto \frac{n_{e}}{H\alpha}$. It is noticed that frequency profiles of both physical quantities look nearly same for same frequency regimes but differ for different frequency regimes.
 \begin{figure}[ht]
\center
\includegraphics[width=240pt,height=200pt]{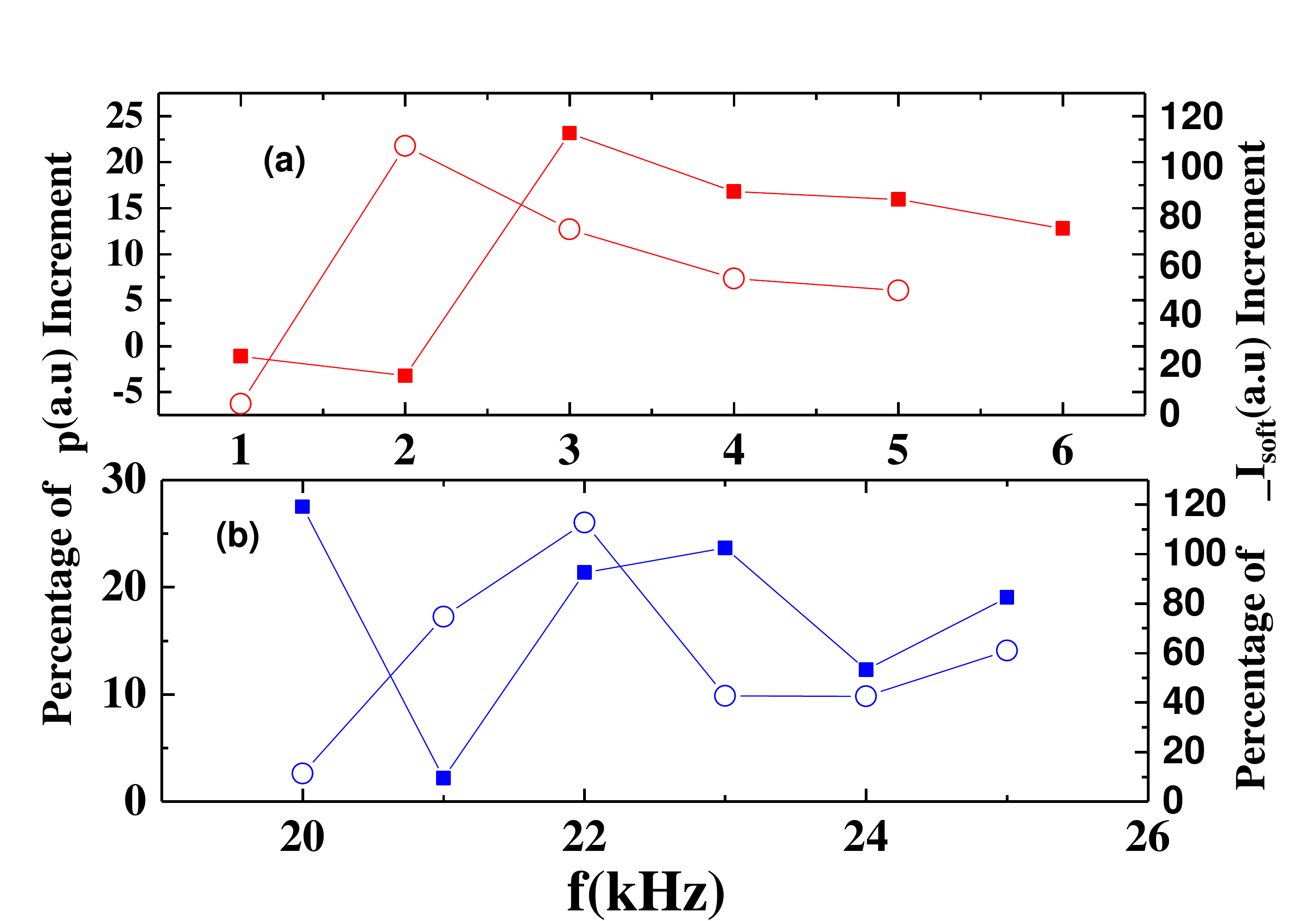}
\caption{Percentage increment of $\tau_{p}$ (solid square box) \& $\tau_{E}$ (hollow circle) (a) lower frequency branch (red line), (b) higher frequency branch(blue line).}\label{fig:7}
\end{figure}
 \par $\tau_{E}$ has also been derived from $\tau_{E\_}kinetic \approx \frac{<n_{e}><T_{e}>v_{plasma}}{I_{P}V_{L}}$. The percentage increment of $\tau_{E\_}I_{soft}$ \& $\tau_{E\_}kinetic$ have been plotted in figure 8(a) \& 8(b) for lower and higher frequency regimes respectively. Interestingly, frequency profile of $\tau_{E\_}I_{soft}$ \& $\tau_{E\_}kinetic$ at low frequency regime look the same and have the peak value at the same frequency (2kHz). On the other hand, $\tau_{E\_}kinetic$ frequency profile is oscillatory which is unlike the $\tau_{E\_}I_{soft}$ frequency profile, clearly shown in figure 8(b). It needs to be mentioned that energy confinement time calculations from $\tau_{E\_}kinetic$ (conventional method) and $\tau_{E\_}I_{soft}$ (newly definition) predict the same frequency peak but their magnitudes are different of the order of 3-5 times.

%\begin{figure}[ht]
%\center
%\includegraphics[width=240pt,height=200pt]{fig6.pdf}
%\caption{Image plot soft x-ray signal (a) without, (b) with biasing.}\label{fig:6}
%\end{figure}

\begin{figure}[ht]
\center
\includegraphics[width=240pt,height=200pt]{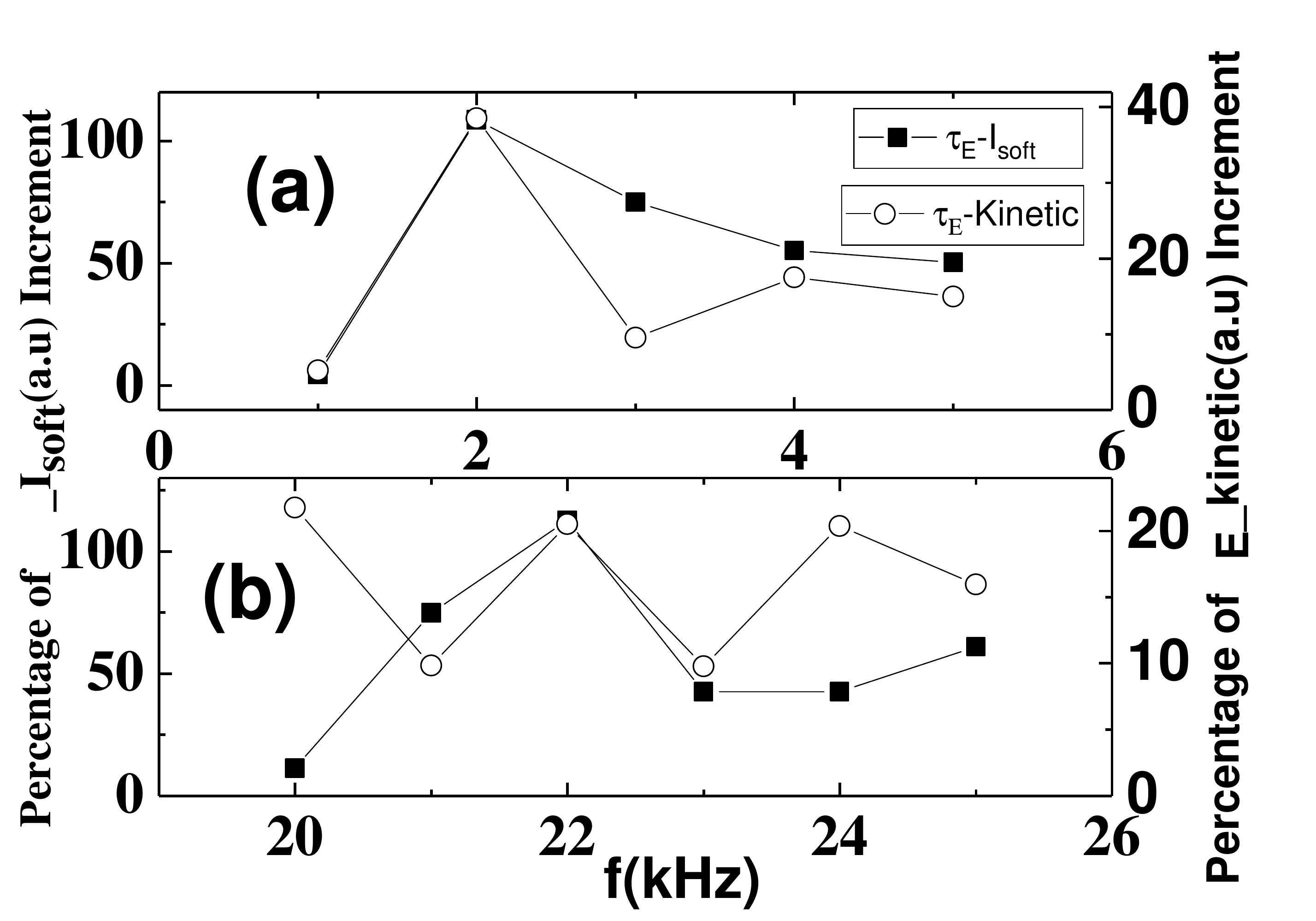}
\caption{Comparison between percentage increment of $\tau_{E\_}I_{Soft}$(solid square box) \& $\tau_{E\_}kinetic$ (hollow circle) (a) lower frequency branch, (b) higher frequency branch.}\label{fig:8}
\end{figure}

\par
In order to find out inherent physics of improved confinement, possible mechanisms related to the effect of AC biased electrode have been explored. Since exciting frequency of applied electric potential remain `kHz' range, therefore formation of zero frequency zonal flow or other mechanisms like ponderomotive force \cite{martinell13} effective at ICR frequencies are not applicable which may stabilize turbulent modes. A possible candidate in relevant frequency range is GAM which can be driven through resonance from oscillating external currents \cite{hung,kn:shur} that create time dependent poloidal flow shear. This produces turbulence suppression by the stochastic Doppler shift due to chaotic stretching of eddies \cite{kn:Diamond}. Accordingly, the fluctuation level is reduced by a factor $(1+ \tau_c \tau_{c,G}\langle k_\theta \tilde V_G^2\rangle)^{-1}$ and it applies for turbulent diffusion coefficient. It is then plausible to expect that high frequency biasing can excite GAM which do not damp as they would if not resonantly excited and it produce some sort of transport barrier that improves confinement. The idea was tested doing a simulation of the plasma evolution using `ASTRA' transport code with this modified turbulent transport coefficients. It \cite{perev} has been run using parameters of STOR-M tokamak in Ohmic heating regime where transport coefficients contain neoclassical as well as turbulent components. Drift wave turbulence can be affected by high-frequency biasing potential in above described mechanisms. Resonant electric field, is produced by electrode confined in a small radial region where turbulence reduction takes place. This would be effective when driving frequency matches with GAM frequency in this edge region of STOR-M which is of the order of $f_{GAM}\sim c_s/R\sim $22 kHz when $T_e\sim $40 eV. The radial electric field is of the order $E_r\sim V/\Delta_r \sim 6000$ V/m, taking potential variation scale length of $\Delta_r\sim 1 $cm and reduced anomalous diffusion coefficient for STOR-M by a factor is given by \cite{martinell13}
\begin{equation}
D=D_0\left( 1+ 0.0012{E_r}^{2/3} f_G \right)^{-1}
\label{eq:trsu}
\end{equation}
where $D_0$ is the diffusion coefficient in absence of oscillating field and $f_G=\tau_c/\tau_{c,G}$, the ratio of correlation times for fluctuations and GAM. Here, $\tau_c=k_\perp^{-2/3}D^{-1/3} S_{v}^{2/3}$ was taken, with $k_\perp$ the radial mode extension assumed as 1 cm, $S_v\sim V/\Delta_r^2 B \sim 10^6 s^{-1}$ the velocity shear. The equilibrium diffusion coefficient is estimated as $D\sim 1\ m^2/s$. The parameter $f_G$ is not known and is taken in the range $1-5$. However, the GAM effect alone is not able to reproduce complete phenomena, so another effect was included. This may come from an increased pinch velocity that confines particles against diffusion. Electrons traveling toroidally parallel to the magnetic field can be thrust by the electrode voltage every time when they passes through its region if the transit time matches the AC bias frequency. This allows the electrons to build a toroidal current which can give rise to a Ware-like radial pinch. It turns out that rotation frequency at the drag velocity for the edge collisionality, $f_{tr} =eE/2\pi R m_e \nu_{ce}$, which extract $\sim 20$ kHz.
\par Both GAM and radial pinch are resonantly driven by AC voltage, the former producing turbulence and transport reduction while the later advects particles to the center. GAM is due to the oscillating radial electric field at the electrode that produces oscillating poloidal flow which reaches steady state at the driving frequency. As a results, transport barrier is formed around electrode.
\par
These elements were incorporated in transport simulations considering different conditions for $f_G$. The results having a better agreement with experimental data correspond to choosing $f_G=3$ for particle diffusion while $f_G=5$ for thermal diffusivity. In this case there is an improvement of plasma confinement in the density while the electron temperature profile becomes less peaked. The results can be seen in figure 9, for a simulation of which the Ware pinch that is initially present is made to increase by 12 \% during biasing.
The simulations were started by running the code without ac biasing until steady state was reached, having plasma parameters in agreement with typical STOR-M values and then the biasing was turned on. The
evolution of representative plasma parameters given by Astra simulation can be observed
in which the high-frequency potential of 22kHz with 60V is applied at time 0.08 s. It is clear
that the behavior agrees in general terms with the discharge shown in figure 3.
The average electron temperature $\langle T_e\rangle$, radiated power from Bremsstrahlung which would be detected in soft X-rays, average and central densities $\langle n_e\rangle, n_{e0}$ and $\tau_{E}$ are increased. Interestingly, particle flux to the
wall which would be related to the $H_\alpha$ emission, shows very small changes which also agrees
with the experimental outcomes. All these trends from simulation support the scenario
of transport reduction by mode stabilization due to resonantly excited GAM associated with
high-frequency biasing potential.
\begin{figure}[t]
\center
\includegraphics[width=200pt]{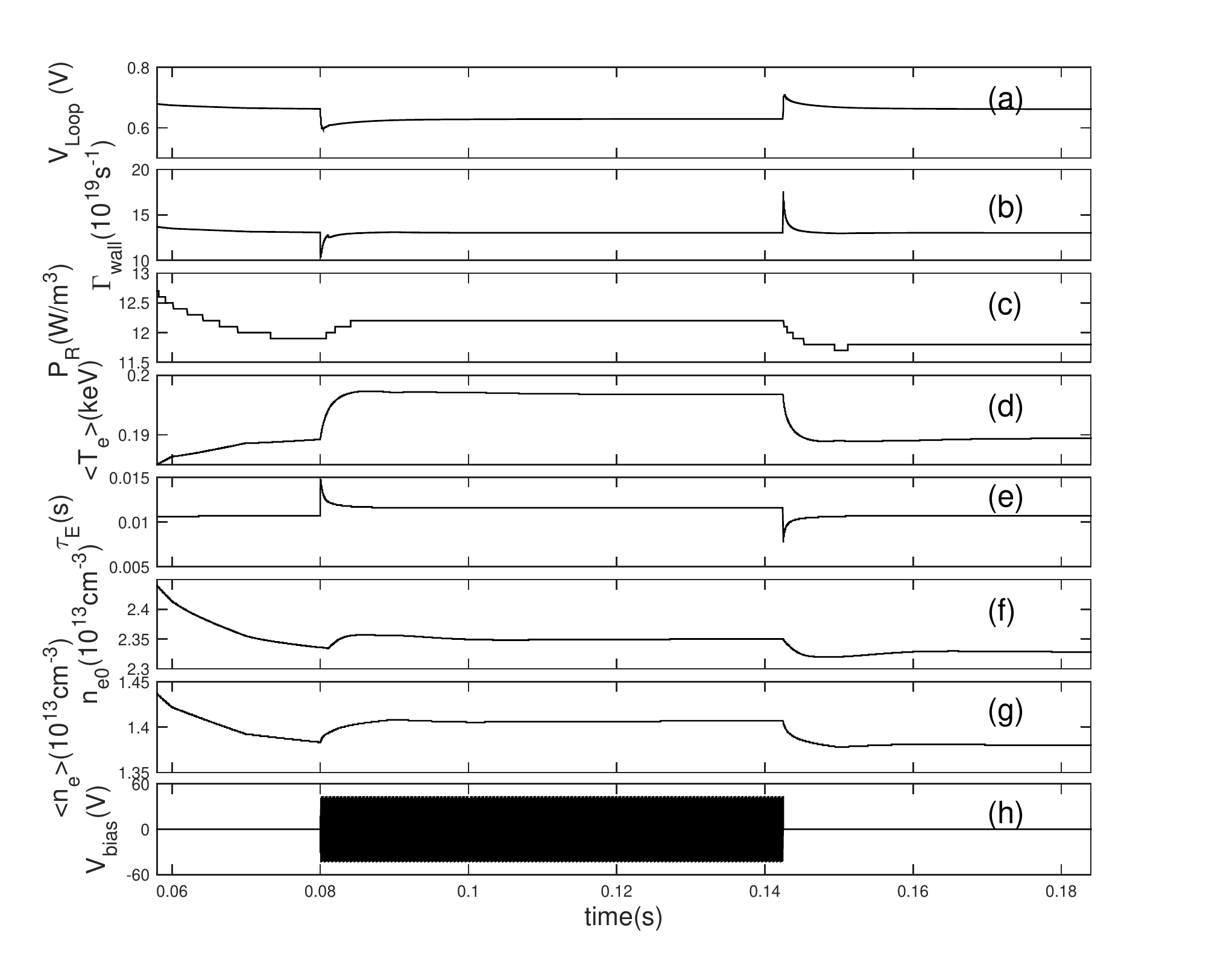}
\caption{Simulation time evolution of STOR-M discharge of (a) loop voltage, (b) particle flux to wall, (c) radiated power, (d) average electron temperature, (e) energy confinement time, (f) central plasma density, (g) average plasma density, (h) high frequency AC bias voltage of 22 kHz.}\label{fig:astra_t}
\end{figure}
\par  According to our knowledge, at first time experimental studies of high frequency biasing in `kHz' range have been performed in tokamak.
Previously, elaborate experimental studies of DC biasing through electrode and limiter \cite{kn:Xiao,kn:zn} were performed in STOR-M. Those experiments show that improved confinement happens at +150V \& -350V with electrode current $\sim$ 22A. The present high frequency biasing experiments show that improved confinement can be achieved with triangular pulse $\pm 60V$ with bias current -24A to +2A at 1kHz-5kHz \& 20kHz-25kHz frequency regimes. The kinetic $\tau_{E}$ increase by $40\%$  and $20\%$ for lower \& higher frequency branches respectively. $\tau_{p}$ is also enhanced around $25\%$ in both frequency regimes. Also, soft X ray emission is enhanced in central part of plasma column, indication of central energy density enhancement. The simulation of proposed model also shows same trend of observed results at higher frequency regimes(20-25)kHz. So, key physics insight for improved confinement at higher frequency regimes may be due to resonant GAM excitation combined with a radial (Ware) pinch produced by the toroidal transit electron current resonantly at bias frequency around 22 kHz. The improved confinement at low frequency regime may be related to sub-harmonic of toroidal transit time. Fluctuations studies and flow measurements will be studied in future.

\par
\begin{acknowledgments}
 We would like to acknowledge NSERC for supporting this work. We also would like to acknowledge machine workshop. Specially, thanks go to Mr. Chomyshen and Mr. Toporowski in the machine workshop for their kind help and friendly approach when needed. The work of A.V. Melnikov was supported by  Russian Science Foundation, project no. 19-12-00312 and in part by the Competitiveness Programme of the National Research Nuclear University ``MEPhI".
\end{acknowledgments}


\section{References:}
\begin{thebibliography}{150}
\bibitem{kn:vanoost1}G. Van Oost et.al., Plasma Phys. Control. Fusion \textbf{45}, 621 (2003)
\bibitem{kn:Ritz}Ch. P. Ritz, R. D. Bengtson, et.al., Phys. Fluids \textbf{27}, 2956 (1984)
\bibitem{kn:Liewer}P. C. Liewer, J. M. McChesney, et.al., Phys. Fluids \textbf{29}, 309 (1986)
\bibitem{kn:Diamond}P.H. Diamond, et.al., Plasma Phys. Control Fusion \textbf{47}, R35 (2005)
\bibitem{kn:deb1}Debjyoti Basu, et.al., Plasma Phys. Control Fusion \textbf{58}, 024001 (2018)
\bibitem{kn:shur}R. V. Shurygin and A. V. Melnikov, Plasma Physics Reports \textbf{44}, 303(2018)
\bibitem{kn:Taylor}R. J. Taylor, et.al., Phys. Rev. Letter \textbf{63} 2365 (1989)
\bibitem{kn:Weynants}R. R. Weynants, et.al., Nucl. Fusion \textbf{32} 837 (1992)
\bibitem{kn:deb}Debjyoti Basu, et.al., Physics of Plasmas \textbf{19} 072510 (2012)
\bibitem{kn:Melnikov} A.V. Melnikov,et.al., Fusion Science and Technology \textbf{46}, 299(2004)
\bibitem{kn:Xiao}C. Xiao et al., Physics of Plasmas \textbf{1}, 2291 (1994)
\bibitem{kn:silva}C. Silva et al., Plasma Phys. Control. Fusion \textbf{46} 163 (2004)
\bibitem{kn:Kan}Zhai Kan et al., Physical Review E \textbf{55} 3431 (1997)
\bibitem{kn:UCKAN}T. UCKAN et al., Nucl. Fusion \textbf{35}, 487 (1995)
\bibitem{kn:Brooks}J. W. Brooks, et.al., Review of Scientific Instruments \textbf{90}, 023503 (2019)
\bibitem{kn:Xiao1}C. Xiao, et.al., Review of Scientific Instruments \textbf{79}, 10E926 (2008)
\bibitem{kn:Heikkinen}J. A. Heikkinen, Physics of Plasmas \textbf{8} 2824 (2001)
\bibitem{kn:Hidalgo}C. Hidalgo, et.al., Plasma Phys. Control. Fusion \textbf{46} 287 (2004)
\bibitem{kn:silver}E.H. Silver et al., Review of Scientific Instruments \textbf{53} 1198 (1982)
\bibitem{hung}C.P. Hung and A.B. Hassam,  Physics of Plasmas \textbf{20}, 092107 (2013)
\bibitem{perev} G. V. Pereverzev, P. N. Yushmanov,  ASTRA: Automated System for TRansport Analysis (Max-Plank-Institute für Plasmaphysik Rep IPP vol 5/98) (Garching: IPP) (2002)
\bibitem{martinell13} J.J. Martinell, et al., Radiation Effects and Defects in Solids \textbf{168}, 866 (2013)
\bibitem{kn:zn} W. Zhang, et al., Phys. Fluids B \textbf{4}, 3277 (1992)
\end{thebibliography}
\end{document}